%% file: article_V4_for_arXiv.tex
\documentclass[12pt,preprint]{aastex}

\usepackage{lineno}
\usepackage{natbib}
\usepackage{times}
\usepackage{subfigure} 
\slugcomment{Version: \today}
\linenumbers
\def\F{{\it Fermi}-LAT }
\def\SU{{\it Suzaku} } 

\input{colordvi.tex}
\usepackage{color}

\shorttitle{X-ray counterparts of high-z blazar candidates}
\shortauthors{Takahashi et al.}  

\title{X-ray and Radio Follow-up Observations of High-Redshift Blazar Candidates in the \F Unassociated Source Population}

\author{Y. Takahashi\altaffilmark{1}, J. Kataoka\altaffilmark{1}, K. Niinuma\altaffilmark{2}, M. Honma\altaffilmark{3,4}, Y. Inoue\altaffilmark{5}, T. Totani\altaffilmark{6,7}, S. Inoue\altaffilmark{8}, T. Nakamori\altaffilmark{1,9}, K. Maeda\altaffilmark{1}}

\email{s072803523@akane.waseda.jp}

\altaffiltext{1}{Research Institute for Science and Engineering, Waseda University, 3-4-1 Okubo, Shinjuku, Tokyo, 169-8555 Japan} 
\altaffiltext{2}{Department of Physics, Faculty of Science, Yamaguchi University, Yoshida 1677-1, Yamaguchi, Yamaguchi, 753-8512, Japan}
\altaffiltext{3}{Mizusawa VLBI Observatory, National Astronomical Observatory of Japan, 2-21-1 Osawa, Mitaka, Tokyo 181-8588, Japan}
\altaffiltext{4}{Department of Astronomical Science, The Graduate University for Advanced Studies, 2-21-1 Osawa, Mitaka, Tokyo 181-8588, Japan}
\altaffiltext{5}{Kavli Institute for Particle Astrophysics and Cosmology, Department of Physics and SLAC National Accelerator Laboratory, Stanford University, Stanford, CA 94305, USA}
\altaffiltext{6}{Department of Astronomy, Kyoto University, Sakyo-ku, Kyoto 606-8502, JAPAN}
\altaffiltext{7}{Department of Astronomy, The University of Tokyo, Bunkyo-ku, Tokyo 113-0033, JAPAN}
\altaffiltext{8}{Max-Planck-Institut f$\ddot{\rm u}$r Kernphysik, Saupfercheckweg 1, 69117 Heidelberg, Germany}
\altaffiltext{9}{Department of Physics, Faculty of Science, Yamagata University, 990-8560 1-4-12 Kojirakawa, Yamagata, Yamagata, Japan}

\begin{abstract}
We report on the results of X-ray and radio follow-up observations of two GeV gamma-ray sources 2FGL~J0923.5+1508 and 2FGL~J1502.1+5548, selected as candidates for high-redshift blazars from unassociated sources in the {\it Fermi} Large Area Telescope Second Source Catalog. We utilize the Suzaku satellite and the VLBI Exploration of Radio Astrometry (VERA) telescopes for X-ray and radio observations, respectively. For 2FGL~J0923.5+1508, a possible radio counterpart NVSS~J092357+150518 is found at 1.4 GHz from an existing catalog, but we do not detect any X-ray emission from it and derive a flux upper limit $F_{\rm 2-8 keV} <$ 1.37 $\times$ 10$^{-14}$ erg cm$^{-2}$ s$^{-1}$. Radio observations at 6.7 GHz also result in an upper limit of $S_{\rm 6.7 GHz}$ $<$ 19 mJy, implying a steep radio spectrum that is not expected for a blazar. On the other hand, we detect X-rays from NVSS~J150229+555204, the potential 1.4 GHz radio counterpart of 2FGL~J1502.1+5548. The X-ray spectrum can be fitted with an absorbed power-law model with a photon index $\gamma$~=~1.8$^{+0.3}_{-0.2}$ and the unabsorbed flux is $F_{\rm 2-8 keV}$~=~4.3$^{+1.1}_{-1.0}$~$\times$~10$^{-14}$~erg~cm$^{-2}$~s$^{-1}$. Moreover, we detect unresolved radio emission at 6.7~GHz with flux $S_{\rm 6.7 GHz}$~=~30.1 mJy, indicating a compact, flat-spectrum radio source. If NVSS~J150229+555204 is indeed associated with 2FGL~J1502.1+5548, we find that its multiwavelength spectrum is consistent with a blazar at redshift $z \sim 3-4$.
\end{abstract}


\begin{document}
\section{INTRODUCTION}
\label{sec:intro}
Blazars, a subclass of active galactic nuclei (AGN) with relativistic jets whose beamed emission is seen within a small angle to our line of sight, are one of the most extreme types of gamma-ray emitting objects. The Large Area Telescope \citep[LAT;][]{atw09} onboard the {\it Fermi} Gamma-ray Space Telescope has detected GeV gamma-ray emission from 781 blazars, whose redshifts have been identified up to z~=~3.214 \citep{2FGL,ack11}. This is a dramatic increase compared to the Third EGRET Catalog of High-Energy Gamma-Ray Sources \citep[3EG catalog;][]{har99} that contained 66 high-confidence identifications of blazars. The number of blazars with identified redshifts z $\geq$ 2 has also increased from seven in the 3EG catalog to 43 in the 2FGL catalog, and three of them have redshifts z $>$ 3. The most distant blazar detected by \F to date is PKS 1402+044 (2FGL~J1405.1+0405) with a redshift z~=~3.214, while the most distant blazar currently known is Q0906+6930 \citep{rom04,rom06} with a redshift z~=~5.47.

Searching for distant blazars is important because 1) their detection would contribute to our further understanding of the cosmological evolution of blazars and their host supermassive black holes \citep[e.g.][]{vol11}, and 2) they can serve as valuable beacons for probing intergalactic environments in the early Universe. Gamma-rays from distant sources can be absorbed through two-photon pair production interactions with softer photons of the extragalactic background light (EBL) \citep[e.g.][; and references therein]{ste06,fra08,ino12}. By analyzing their gamma-ray spectra, we can constrain or potentially measure the gamma-ray opacity and the EBL in a redshift-dependent way \citep[see e.g.][]{abd10,orr11,ack12a,abr13}. Furthermore, we may obtain unique insight into the cosmic reionization epoch with gamma-ray sources at $z \gtrsim 6$ \citep{oh01,ino10,ino12}.

Employing a model of the blazar gamma-ray luminosity function based on EGRET observations, the X-ray luminosity function of general active galactic nuclei (AGN) and the optical luminosity function of quasars by the Sloan Digital Sky Survey (SDSS),
\citet{ino11} showed that \F may be able to detect blazars with redshifts up to z~=~5$-$6 after a five-year survey, assuming a corresponding \F flux detection limit of F($>$ 100 MeV)~=~1$\times$10$^{-9}$ photons cm$^{-2}$ s$^{-1}$. Such distant sources are expected to be faint with fluxes comparable to the detection limit. During the first two years of operation, \F has detected many faint gamma-ray sources, but most of them remain unassociated with known classes of astronomical objects. We speculate that some of them may indeed be blazars with intrinsically high luminosities but high redshifts, z $\gtrsim$ 3.

The spectral energy distributions (SEDs) of luminous blazars typically consist of two, broadly peaked components. The low energy peak extending from the radio to optical/UV bands is understood as synchrotron emission from relativistic electrons or positrons,
while the high energy peak covering the X-ray and gamma-ray bands is widely believed to be primarily inverse Compton emission.
In order to identify unassociated gamma-ray sources with blazars, multiwavelength characterization of their SEDs,
particularly of the above two components, is essential. Since there already exist several deep radio and optical surveys that cover a large fraction of the sky, we can utilize them for the purpose of clarifying the SEDs. On the other hand, in X-rays, deep observations with sensitivities reaching $\sim 10^{-14}$~erg~cm$^{-2}$~s$^{-1}$ are limited to pointing observations, which have been carried out only for the brighter unassociated \F sources. Therefore, we conducted new X-ray observations using the \SU satellite. Observing in X-rays also has the merit of potentially discovering high-energy variability, a crucial characteristic of AGNs. We also carried out radio observations at 6.7 GHz frequency, higher than in available catalogs, to clarify the radio spectra using the VLBI Exploration of Radio Astrometry telescopes \citep[VERA; ][]{kob05}.

In this paper, we report on the results of our radio and X-ray observations of two radio sources that are selected as counterparts of possible distant gamma-ray blazars. In the following section, we outline criteria for selecting distant blazars from \F unassociated sources and apply them to the 2FGL catalog. Then we describe the details of our radio and X-ray observations, data reduction, and analysis procedure. We present the results of our observations in Section~\ref{sec:results}. Finally, we discuss the properties of the gamma-ray sources based on the available multiwavelength information.

\section{OBSERVATIONS AND DATA ANALYSIS}
\label{sec:obs-anal}

\subsection{Source Selection}
\label{sec:src-sel}

In the 2FGL catalog, detection of point sources involves iterating 
through three steps as described in detail in \citet{2FGL}: 
(1) identification of potential point sources, denoted as ``seeds''; 
(2) a full all-sky optimization of a model of the $\gamma$-ray
sky including the new seeds to refine their estimated positions and 
evaluate their significances; and (3) creation of a ``residual test 
statistic (TS) map.'' The TS is evaluated as TS = 2(log $\mathcal{L}$(source) 
$-$ log $\mathcal{L}$(nosource)), where $\mathcal{L}$ represents the likelihood of the data given the model with or without a source present at a given 
position on the sky. To evaluate the fluxes and spectral parameters, 
the sky was split into 933 Regions of Interest (RoI) in order to make 
the log $\mathcal{L}$  maximization tractable.
The source photon fluxes are reported in the 2FGL catalog in 
five energy bands (100--300 MeV; 300 MeV to 1 GeV; 1--3 GeV; 3--10 GeV; 
10--100 GeV). The fluxes were obtained by freezing the 
spectral index to that obtained in the fit over the full range and 
adjusting the normalization in each spectral band. 
For bands where the source was too weak to be detected, 
those with TS in the band TS$_{i}$ $<$ 10 or relative uncertainty 
on the flux $\Delta$ $F_i$/$F_i$ $>$ 0.5, 2$\sigma$ upper limits 
were calculated, $F_i^{\rm UL}$.

In order to select candidates for distant gamma-ray blazars from \F unassociated sources, \citet{ino11} suggested source selection criteria based on expected multiwavelength spectral features. In this paper, we apply these criteria to actual \F unassociated sources from the 2FGL catalog with some small modification, as summarized below. First of all, since high-redshift blazars are naturally expected to be faint and variable, we select faint gamma-ray sources with flux $\simeq$~10$^{-10}$ photon cm$^{-2}$ s$^{-1}$ in the 1$-$100 GeV band, as well as with significant variability, identified at 99 \% confidence with the relation TS$_{\rm var}$ $\geq$ 41.64 in terms of the variability index TS$_{\rm var}$ as defined in Eq. (4) of \citet{2FGL}. Second, the sources should have soft gamma-ray spectra with power-law photon indices $\Gamma > 2.3$ as measured through spectral fitting in the 100 MeV $-$ 100 GeV range, in accord with the ``blazar sequence'' \citep{fos98,kub98}, the observed tendency for the SEDs of more luminous blazars to have lower peak frequencies for the synchrotron and inverse Compton components, despite ongoing debate as to whether the blazar sequence is an intrinsic physical property of blazars or simply due to observational biases \citep{pad07,gio12}. Third, the sources should additionally have a compact radio counterpart with intensity $\geq$ 20 mJy. Finally, they should either lack an optical counterpart or show evidence of a Lyman break in the optical band due to absorption by intergalactic neutral hydrogen.

To search for gamma-ray sources that meet the above criteria, we need radio and optical catalogs covering a large fraction of the sky. The optical catalog should also have data in multiple color bands in order to allow examination of Lyman breaks. For this purpose we utilized the Eighth Data Release of the Sloan Digital Sky Survey \citep[SDSS catalog;][]{ade11}, for which data are available in five colors. Therefore our study is limited to the gamma-ray sources in regions of the sky with SDSS coverage. For radio counterpart searches, we used the NRAO VLA Sky Survey \citep[NVSS;][]{con98} with sky coverage of b $>$ $-40^{\circ}$ at 1.4 GHz frequency. Additionally, we examined only gamma-ray sources at high Galactic latitudes ($|$b$|$ $>$ $10^{\circ}$) so as to avoid contamination by Galactic sources.

After the selection, two unassociated \F sources remain, 2FGL~J0923.5+1508  with TS$_{\rm var} = 60.36$ and $\Gamma = 2.33$, and 2FGL~J1502.1+5548 with TS$_{\rm var} = 46.61$ and $\Gamma = 2.65$. Each has a candidate radio counterpart detected at some frequency. All the radio counterparts collected from the available radio catalogs are listed in Table~\ref{tab:radio_ctp}. NVSS~J092357+150518 is a possible 1.4 GHz counterpart of 2FGL~J0923.5+1508 and we could not find any optical counterpart in the SDSS catalog. NVSS~J150229+555204 is a candidate 1.4 GHz counterpart of 2FGL~J1502.1+5548. For this radio source, we find an optical counterpart with evidence of intergalactic attenuation in the u band in the SDSS catalog, SDSS~J150229.04+555205.2 with magnitudes u $>$ 22.3, g~=~19.80, r~=~19.35, i~=~19.06, and z $>$ 20.8, where we quote 5$\sigma$ upper limits for the u and z bands \citep{lve00}. Central wavelengths for each color band are 3551~\AA, 4686~\AA, 6166~\AA, 7480~\AA, and 8932~\AA\, for u, g, r, i and z, respectively. The non-detection in the z band may be due to saturation by the nearby bright star SDSS~J150228.45+555209.4 (with magnitudes u~=~13.62, g~=~11.98, r~=~11.40, i~=~11.22, z~=~11.16), in addition to the low sensitivity in the z band. If we assume that the Lyman break lies in between the u and g bands, the redshift of this optical counterpart should be $\sim 3-4$. Both radio sources are the brightest in the \F error region. Since the radio counterpart did not have X-ray counterparts in any available archival X-ray catalogs, deep X-ray observations were required to study the multiwavelength properties of these sources. Thus we utilized the \SU satellite for this purpose. Additionally, we conducted new radio observations at frequencies higher than 1.4 GHz using VERA to clarify the spectrum and morphology of the radio emission. The two gamma-ray sources are listed in Table\,\ref{tab:obs_log} together with some relevant parameters from the 2FGL catalog and \SU observation logs.

\subsection{\SU Observations and Data Reduction}
\label{sec:obs-reduct}
The observations were conducted with the three X-ray Imaging Spectrometers \citep[XIS;][]{koy07} and the Hard X-ray Detector \citep[HXD;][]{kok07,tak07}. The XIS detectors are composed of four CCD cameras, one of which (XIS1) is back-illuminated and the others (XIS0, XIS2, and XIS3) front-illuminated. The operation of XIS2 ceased in 2006 November because of contamination by a leaked charge. Since none of the studied sources have been detected with the HXD, below we describe the analysis of only the XIS data. The XIS were operated in the pointing mode and the normal clocking mode, combined with the two editing modes 3 $\times$ 3 and 5 $\times$ 5.

We conducted all the data reduction and analysis with \texttt{HEADAS} software version 6.11 and the calibration database (CALDB) released on 2012 February 10. First, we combined the cleaned event data of the two editing modes using \texttt{xselect}. Then we removed the data corresponding to periods when the \SU satellite was passing through the South Atlantic Anomaly (SAA) and up to 60\,sec afterwards, as well epochs of low-Earth elevation angles (less than $5^{\circ}$). We also excluded the data obtained when the \SU satellite was passing through regions of low Cut-Off Rigidity (COR) below 6\,GV. In addition to the above data reduction, the data obtained by XIS1 required removal of events in the rows next to the charge injected rows (second trailing rows), because the increased amount of charge injection has led to an increase in the NXB level since June 2011. Finally, we removed hot and flickering pixels using \texttt{sisclean} \citep{day98}.

\subsection{{\it SUZAKU} DATA ANALYSIS}
\label{sec:sudanalys}
We extracted X-ray images from the two operating front-illuminated CCDs (XIS0 and XIS3). Then, Non X-ray Background (NXB) subtraction and an exposure correction were applied to the extracted images. After that, we combined the X-ray images of XIS0 and XIS3, which were then finally smoothed using a Gaussian function with $\sigma$~=~$0'.28$. The resulting images are presented in Figures \ref{fig:imgJ0923} and \ref{fig:imgJ1502} and discussed further in the section\,\ref{sec:results}. Positional errors of each gamma-ray source taken from the 2FGL catalog and the positions of radio sources corresponding to candidate distant blazars are also shown on each X-ray image with thick green ellipses and green crosses respectively.

For further analysis, we selected source regions around each detected X-ray source within the 2FGL error ellipses. The radii of the source extraction regions were set to $1'$ or $2'$. The corresponding background regions with radii of $3'$ were taken from the low count rate area in the same XIS chips (dashed green circles). We set the detection threshold for X-ray sources at $4\sigma$, based on the signal-to-noise ratio defined as the ratio of the number of excess events above background to its standard deviation assuming a Poisson distribution. The X-ray source positions and the corresponding errors were estimated by 2D Gaussian fits.

Then, we conducted detailed spectral and timing analysis of each detected X-ray source inside the \F error ellipse. For the timing analysis, light curves from the front-illuminated (XIS0, XIS3) and back-illuminated (XIS1) CCDs were summed after subtracting the corresponding backgrounds using \texttt{lcmath}. The light curves constructed in this way provide the net-count rates. To quantify possible flux variations, a $\chi^2$ test was applied to each light curve using \texttt{lcstats}. For the X-ray spectral analysis, we generated the RMF files for the detector response and the ARF files for the effective area using \texttt{xisrmfgen} and \texttt{xissimarfgen} \citep{ish07}. When generating an ARF file of XIS1, we set the option `pixq\_and~=~327680' to remove events in the second trailing rows. In order to improve the statistics, we combined the data from the two front-illuminated CCDs using \texttt{mathpha} without calculating the Poisson errors, and then combined the response files using the \texttt{marfrmf} and \texttt{addrmf} commands. Uncertainties of the model spectral parameters are computed at $90\%$ confidence levels. The results of the timing and spectral analysis of the two targeted sources are summarized in Table\,\ref{tab:fits_target}, and discussed below in more detail.

\subsection{VERA Observations and Data Reduction}
\label{sec:vera}
VERA observations of NVSS J092357+150518 and NVSS J150229+555204 were conducted on 2011 Nov 10 and 11 using three stations of the VERA array. The observations were done at 6.7 GHz, and the typical system noise temperature was $\sim$ 120 K, which was measured every 10 minutes using the chopper-wheel dummy load at room temperature. We recorded the left-handed circular polarization signal at the data rate of 1 Gbps, which provides a total recording bandwidth of 256 MHz with two-bit quantization. Both sources were observed for 40 minutes in total. The correlation processing of the data from the three VERA stations was carried out using the Mitaka FX correlator.

For the correlated visibility, we conducted the standard VLBI calibration using the NRAO AIPS package. The amplitude calibration was carried out based on a priori calibration using the system noise temperature obtained during the observations. In the fringe search procedure, the AIPS task FRING was used. First we searched fringes for the fringe finder sources to calibrate the clock offset and clock rate offset, and then by using these clock parameters, we searched for fringes of NVSS J092357+150518 and NVSS J150229+555204. In order to find fringes of faint sources, we used the following setup for the fringe search process: 1) integration time in the fringe search was set to be 5 minutes, which corresponds roughly to an empirically-determined coherence time of VERA at 6.7 GHz, 2) to reduce the probability of false detection, search windows of delay and rate offsets were set to be $\pm10$ nsec and $\pm10$ mHz, respectively (this window size corresponds to 10 $\times$ 6 independent grids in the delay-rate window), and 3) we set a baseline-based signal-to-noise ratio (SNR) cutoff of 2 as detection threshold, which corresponds roughly to a minimum detection flux of $\sim$ 20 mJy, and station-based fringe solutions were solved from baseline-based fringes beyond this detection threshold. For NVSS J092357+150518, no station-based solutions consistent with baseline-based fringes were obtained, and thus we conclude that this source was not detected. On the other hand, for NVSS J150229+555204 we detected possible fringes from all the eight segments of 5 min data (coming from four 10-min scans). For a consistency check, we have confirmed that delays of the fringes are consistent with each other for all the eight segments. While the baseline-based fringe search was done using low SNR cutoff (SNR of 2 corresponding to 95\% confidence level), the consistency of the delay for the eight different data points assures that the false detection rate is as low as 10$^{-8}$, and thus we conclude that the fringe from NVSS J150229+555204 is a true detection.

\section{RESULTS}
\label{sec:results}
\subsection{NVSS~J092357+150518}
In the observation of NVSS~J092357+150518, we detect an X-ray source located $\sim 1'.6$ away from the position of the NVSS source. We present the X-ray image obtained by the \SU XIS in Figure~\ref{fig:imgJ0923}. However, the pointing uncertainty of \SU is estimated to be $\leq 1'$. Therefore, this X-ray source is unlikely to be an X-ray counterpart of NVSS~J092357+150518. To calculate the X-ray upper limit, we determine a source region and a background region as indicated in Figure\,\ref{fig:imgJ0923} with solid and dashed lines, respectively. Then we calculated a 90\% confidence level upper limit for the X-ray flux of NVSS~J092357+150518 by assuming an absorbed power-law model with fixed parameters $N_{\rm H}$~=~3.51 $\times$ 10$^{20}$ cm$^{-2}$ (derived from \citet{dic90}) and a photon index $\gamma$~=~2.0, resulting in 1.37 $\times$ 10$^{-14}$ erg cm$^{-2}$ s$^{-1}$.

The fringe of NVSS J092357+150518 was not detected by the VERA observation, even though we set a baseline-based SNR cutoff of 2 and wide search windows in the fringe search process. Therefore we conclude that this source is not detected, and estimate a correlated flux upper limit of $S_{6.7GHz}~<~$19.0 mJy at approximately 30 M$\lambda$ at $2\sigma$. Here $1\sigma$ is the noise level, which is derived from the baseline sensitivity of the VERA Mizusawa $-$ Ogasawara baseline. The flux density of the central few milli-arcsec region of this source may be much fainter than the one obtained by previous observations as shown in Table 1, which may be due to an extended source structure that is partially resolved by VLBI.

\subsection{NVSS~J150229+555204}
We present the X-ray image for the observation of NVSS~J150229+555204 in Figure~\ref{fig:imgJ1502}. We detect an X-ray point source with a significance of 11.2$\sigma$ at the position coincident with NVSS~J150229+555204. The position of the detected X-ray counterpart is [R.A., decl.]~=~[225.627(4), 55.872(4)]. For the detailed timing and spectral analyses, we determine extraction regions of background events and X-ray source events as shown in Figure~\ref{fig:imgJ1502}. The light curve of the X-ray counterpart with a time binning of 5760 s and its spectra are presented in Figures~\ref{fig:lcJ1502} and \ref{fig:specJ1502}, respectively. In the timing analysis, the light curve can be fitted with a constant count rate with $\chi^{2}$/d.o.f.~=~11.7/13. In the spectral analysis, we fitted the X-ray spectrum with an absorbed power-law model. The value of $N_{\rm H}$ = 1.46 $\times 10^{20}$ cm$^{-2}$ was fixed as derived in \citet{dic90}. This model provided the best fit with a photon index $\gamma~=~1.8^{+0.3}_{-0.2}$ and $\chi^{2}$/d.o.f.~=~29.5/32. The derived unabsorbed flux in the 2-8 keV energy range is 4.3$^{+1.1}_{-1.0}$ $\times$ 10$^{-14}$ erg~cm$^{-2}$~s$^{-1}$.

In the case of NVSS J150229+555204, we detected the fringes based on the procedure in section~\ref{sec:vera}. The calibrated visibilities were then exported to carry out an imaging procedure using the Caltech Difmap package. As a result, we clarified the compact structure of this source (Figure~\ref{fig:vera_j1502}), and the VLBI flux is $S_{6.7GHz}~=~30.1$ mJy, which is derived from 2D Gaussian fitting to the visibility data in the ({\it u-v})-plane using the task {\it modelfit} in the Difmap.

\subsection{Other Detected X-ray Sources}
In addition to the X-ray counterpart of NVSS~J150229+555204, we also detected multiple X-ray sources inside the 95\% positional error regions of the gamma-ray sources as noted in the 2FGL catalog, although our observation did not cover the entire error regions. In the observation of NVSS~J092357+150518, we discovered four X-ray sources that are not listed in any previous X-ray catalogs. Similarly, in the observation of NVSS~J150229+555204, we discovered two new X-ray point sources and detected two 1RXS \citep[ROSAT All-Sky Survey Source Catalogue; ][]{vog99,vog00} sources (1RXS~J150134.9+555047 and 1RXS~J150220.6+554830), one of which is the cluster of galaxies MHL~J150137.1+555056 \citep{wen12}. X-ray positions of these sources, their detection significance, and the radii of event extraction regions are listed in Table \ref{tab:det_srcs} with source numbers that correspond to those marked in Figure~\ref{fig:imgJ0923} and Figure~\ref{fig:imgJ1502}.

We also analyzed the light curves and spectra of these X-ray sources. The $\chi^{2}$ fit to the light curves assuming constant fluxes resulted in only source 5 to be statistically variable with $\chi^{2}$/d.o.f.~=~42.7/13. In the spectral analysis, we fitted all the spectra with an absorbed power-law model. For sources with moderate absorption, the values of $N_{\rm H}$ were fixed at those derived in \citet{dic90}, and for sources resulting in bad fits, we tried other spectral models and determined the best fit model. The best fit spectral models and parameters are summarized in Table~\ref{tab:fits}. 

For all the X-ray sources we detect inside the 2FGL error regions, we searched for radio, infrared, and optical counterparts from the VLA Faint Images of the Radio Sky at Twenty-cm \citep[FIRST;][]{bec95} Survey, the Wide-field Infrared Survey Explorer \citep[WISE;][]{wri10} All-Sky Source Catalog, and the SDSS catalog, respectively. We take the sources nearest to the X-ray sources as counterparts and summarize them in Table~\ref{tab:counterparts}.

\section{DISCUSSION AND CONCLUSIONS}
\label{sec:DandC}
In this paper, we report on the results of X-ray and radio follow-up observations as well as counterpart searches with existing multiwavelength catalogs for two \F unassociated sources that have been selected as candidate distant blazars based on the criteria described in section~\ref{sec:obs-anal}. For NVSS~J092357+150518, the potential 1.4 GHz counterpart of 2FGL~J0923.5+1508, we do not detect X-ray emission and derive a stringent upper limit to the X-ray flux in the 2-8 keV energy range of $F_{2-8 {\rm keV}} < 1.37 \times 10^{-14}$ erg cm$^{-2}$ s$^{-1}$. The radio observation with VERA at 6.7 GHz also resulted in no detection with an upper limit of S$_{6.7\rm GHz} <$ 19 mJy. In Figure \ref{fig:seds}(a), we present the spectral energy distribution of 2FGL~J0923.5+1508, assuming NVSS~J092357+150518 as the radio counterpart. Combining with non-contemporaneus archival data at 74 and 365 MHz, the radio spectral index is constrained to be $\alpha_{\rm r} \geq$ 0.94 where the radio flux S$_{\nu} \propto \nu^{-\alpha_{\rm r}}$, which is much steeper than typically expected for blazars. Note that although this steep radio spectrum could already be inferred from the archival data alone, our new upper limit at 6.7 GHz significantly strengthens the case. NVSS~J092357+150518 is more likely to be a steep spectrum radio quasar, a subclass of radio loud quasars with $\alpha_{\rm r} >$ 0.5 \citep{lan04}, in which case the \F source may be unrelated.

On the other hand, for NVSS~J150229+555204, the potential 1.4 GHz counterpart of 2FGL~J1502.1+5548, we detect X-rays with a flux $F_{2-8 {\rm keV}}~=~4.3^{+1.1}_{-1.0} \times 10^{-14}$ erg cm$^{-2}$ s$^{-1}$. The photon index is $\gamma$~=~1.8$^{+0.3}_{-0.2}$, typical of AGN \citep{toz06,mat10}, confirming the non-thermal nature of the X-ray emission from the radio counterpart. We also detect radio emission with our new VERA observation at 6.7 GHz with a flux $S_{6.7 {\rm GHz}}~=~30.1$ mJy. In Figure~\ref{fig:seds}(b), we present the spectral energy distribution of 2FGL~J1502.1+5548 when adopting NVSS~J150229+555204 as the radio counterpart. The radio spectral index $\alpha_{\rm r}$ is 0.19$\pm$0.05, consistent with typical values for blazars ($\alpha_{\rm r} < 0.5$). The optical spectral index is $\alpha_{\rm o}$ = 1.35, where $S \propto \nu^{-\alpha_{\rm o}}$. Given its likely identification as a blazar, we can attempt to classify this source from the peak frequency of the synchrotron component in the broad band spectra. Since we do not have observations covering the actual peak frequency, we instead determine broad-band spectral indices $\alpha_{\rm ro}$ (between 5 GHz and 5000 $\AA$) and $\alpha_{\rm ox}$ (between 5000 $\AA$ and 1 keV) where $f_{\rm ro} \propto \nu^{-\alpha_{\rm ro}}$ and $f_{\rm ox} \propto \nu^{-\alpha_{\rm ox}}$ as defined in \citet{ack11}. We estimate the radio and optical fluxes at 5 GHz and 5000 $\AA$ by extrapolating the power laws measured in the radio and optical bands, respectively. As the u-band non-detection of the optical counterpart can be interpreted as a Lyman break for a source at $z \sim 3-4$, we can assign a tentative redshift z~=~3.5 (changing z by $\sim~0.5$ level does not significantly affect the results). Then the rest-frame broad band spectral indices would be $\alpha_{\rm ro}$~=~0.39$\pm 0.01$ and $\alpha_{\rm ox}$~=~1.31$\pm 0.08$. According to Figure~7 of \citet{ack11}, blazars with these values can be either Intermediate Synchrotron Peaked (ISP) blazars or Low Synchrotron Peaked (LSP) blazars. Blazar SED sequence models as discussed in \citet{ino09} and including intergalactic attenuation with the EBL model of \citet{ino12} are also plotted in Figure~\ref{fig:seds}(b). The red dashed, green dot-dashed and blue solid curves represent models with a gamma-ray luminosity $L_{\gamma}~=~10^{47.5}$ erg s$^{-1}$ at 100 MeV and assuming redshifts z~=~3.0, 3.5, 4.0, respectively. They provide a good match to the available observations and the synchrotron peak frequency of the model is consistent with LSP blazars ($\nu_{\rm peak}~\lesssim~10^{14}$~Hz). Although these simplified SED models are seen to overestimate the radio fluxes, a proper account of synchrotron self absorption effects should bring them into better agreement \citep[e.g.,][]{ryb79}.

Based on the strength of their optical emission lines, blazars can also be classified into BL Lacertae objects (BL Lacs) with weak or no lines, or flat spectrum radio quasars (FSRQs) with strong lines. The gamma-ray luminosity of these two classes are known to be systematically different, with that of FSRQs being higher and also showing a higher ratio relative to the synchrotron luminosity compared to BL Lacs. These facts have been interpreted in terms of emission models where the synchrotron self Compton \citep[SSC,][]{mar92,blo96,tav98} and external Compton \citep[EC, e.g.][]{sik94, der02} processes contribute to the gamma-rays at different levels for each class \citep{ino96,ghi09,ghi10}. Blazars that are detectable to higher redshifts are more likely to be FSRQs in view of their higher luminosities. As our purpose is to find the most distant blazars, associations of our target sources with FSRQs will reinforce our case. In Figure~\ref{fig:a-gpl}, we plot $\alpha_{\rm ro}$ versus gamma-ray photon indices $\Gamma$ of BL Lacs (dark blue for HSP, light blue for ISP, and green for LSP) and FSRQs (red filled circles) listed in the Second Catalog of Active Galactic Nuclei Detected by the Fermi Large Area Telescope \citep[2LAC;][]{ack11}, compared with those for 2FGL~J1502.1+5548 (a black star). In the $\alpha_{\rm ro}$-$\Gamma$ plane, each blazar class can be differentiated. Most blazars with spectral parameters in the upper right area of this plane are FSRQs, which are intrinsically bright. According to the blazar sequence, such objects would have large $\Gamma$ and synchrotron peaks between the radio and optical bands, leading to relatively large $\alpha_{\rm ro}$. On the other hand, blazars with parameters in the lower left region are BL Lacs that have a wide range of $\Gamma$ and relatively small $\alpha_{\rm ro}$. Figure~\ref{fig:a-gpl} indicates that the spectral properties of 2FGL~J1502.1+5548 are consistent with an FSRQ, although an extreme ISP/LSP BL Lac may also be possible.

Finally, we discuss the possibility that some of the other sources detected in X-rays besides our targeted radio sources are in fact the gamma-ray emitters. In both observations, we detected multiple X-ray sources inside the positional error boxes of 2FGL~J0923.5+1508 and 2FGL~J1502.1+5548. In the case of 2FGL~J0923.5+1508, we detected four X-ray point sources and all of their spectra were fitted well with absorbed power-law models with photon indices 1.4-2.0. Taking into account the relatively large statistical errors, the spectra of these four sources are consistent with AGNs. Since all lack radio counterparts, they are probably radio-quiet AGNs. Theoretical studies suggest that radio-quiet AGNs may emit gamma-rays by the decay of neutral pions produced in a hot accretion flow near the black holes \citep{oka03} or Comptonization by non-thermal electrons in coronae above accretion disks \citep{ino08}. However, gamma-ray emission from radio-quiet AGNs have not been confirmed even from bright hard X-ray selected Seyfert galaxies, except for some type-2 Seyferts with starburst activity \citep{ten11,ack12b, ack12c}. The four sources here are two orders of magnitude fainter in X-rays than the Fermi-detected Seyferts, and moreover, gamma-ray emission of starburst origin is expected to be accompanied by detectable radio and/or optical emission. Thus, they are unlikely to be associated with 2FGL J0923.5+1508.

In the case of 2FGL~J1502.1+5548, there are two sources with X-ray spectra that can be fitted with absorbed power-law models and have photon indices consistent with AGNs. One of these two (source 7) has a radio counterpart with flux 7.89 mJy at 1.4 GHz. However, this flux is below the threshold of our criteria 20 mJy and likely too faint to be detected by \F. Therefore, together with the above discussion about radio-quiet AGNs, these two sources do not appear to be X-ray counterparts of the gamma-ray source. There is a cluster of galaxies MHL~J150137.1+555056 inside the error region of 2FGL~J1502.1+5548. 
Although radio-loud galaxies that are members of clusters have been detected in gamma rays~\citep{abd09a,abd09b},
clusters themselves are yet to be detected. Lacking bright radio counterparts, MHL~J150137.1+555056 is unlikely to be a gamma-ray emitter. The brightest variable X-ray source detected inside the error region of 2FGL~J1502.1+5548 is source 5 that has a spectrum fitted well with an absorbed power-law model combined with an APEC model for emission from optically thing thermal plasma \citep{smi01}. Although this implies that the source is an active high energy object, the gamma-ray emission cannot lie on a simple extension of the X-ray spectrum with its relatively soft photon index $\gamma~=~2.33^{+0.13}_{0.13}$.

From our X-ray and radio observations, we found that 2FGL~J1502.1+5548 is highly likely to be a distant gamma-ray emitting blazar. To determine the exact redshift of this source, detailed optical spectroscopy with large telescopes is required. Further multiwavelength studies with current and future instruments are desirable in order to clarify its blazar classification, including deeper X-ray observations with {\it XMM-Newton} or {\it Chandra}, hard X-ray observations with the {\it Nuclear Spectroscopic Telescope Array} and {\it ASTRO-H}, deeper gamma-ray observations above $\sim$30 GeV with the {\it Cherenkov Telescope Array}, and sub-millimeter observations with the {\it Atacama Large Millimeter Array}.

\input{ref_v2.tex}
\input{tables_v2.tex}

\input{figures_v2.tex}

\end{document}

%% file: tables_v2.tex
\begin{table}[m]
\small
\caption{Radio Counterparts of Gamma-ray Sources Selected as Candidate High Redshift Blazars}
\label{tab:radio_ctp}
\begin{center}
\begin{tabular}{clcc}
\hline\hline
2FGL Name & Radio Counterpart & Observed Frequency & Radio Flux (mJy)\\ 
\hline
2FGL J0923.5$+$1508 & VLSS~J0923.9+1505   & 74 MHz  & 1580 \\
                    & TXS~0921+153        & 365 MHz & 373 \\
                    & NVSS~J092357+150518 & 1.4 GHz & 70.5 \\
2FGL J1502.1$+$5548 & WN 1501.0+5603      & 325 MHz & 50.0 \\
                    & NVSS~J150229+555204 & 1.4 GHz & 34.8 \\
                    & GB6 J1502+5552      & 4.85 GHz & 24.1 \\
\hline
\end{tabular}
\end{center}
Notes. Counterparts with corresponding identifiers were found from the following catalogs. VLSS; VLA Low-Frequency Sky Survey~\citep{coh07}, TXS; Texas Survey of Radio Sources~\citep{dou96}, WN; Westerbork Northern Sky Survey~\citep{ren97}, and GB6; Green Bank 6 cm Radio Source Catalog~\citep{gre96}.
\end{table}

\begin{table}[m]
\scriptsize
\caption{Gamma-ray properties and Suzaku observation logs of the studied sources}
\label{tab:obs_log}
\begin{center}
\begin{tabular}{cccccccc}
\hline\hline
Name & TS$_{\rm var}$$^{\rm a}$ & $\Gamma^{\rm b}$ & OBS ID & \multicolumn{2}{c}{Pointing Direction$^{\rm c}$} & Observation start &
 Effective exposure \\ 
   & & & & \multicolumn{1}{c}{RA [deg]} & \multicolumn{1}{c}{DEC [deg]} & (UT) & [ksec] \\
\hline
2FGL J0923.5$+$1508 & 60.36 & 2.33 & 707007010 & 140.9890 & 15.0880 & 2012/04/29 06:20:35 & 86.7\\
(NVSS~J092357+150518)& & & & & & & \\
2FGL J1502.1$+$5548 & 46.61 & 2.65 & 707008010 & 225.6210 & 55.8680 & 2012/05/22 21:54:41 & 53.5\\
(NVSS~J150229+555204)& & & & & & & \\
\hline
\end{tabular}
\end{center}
$^{\rm a}$ Variability index (for more detail, see 2FGL catalog \citep{2FGL})\\
$^{\rm b}$ Power-law photon index in the 2FGL catalog, where dN/dE $\propto$ E$^{-\Gamma}$\\
$^{\rm c}$ Planned target coodinates taken from the positions of the NVSS counterparts.
\end{table}

\begin{table}[m]
\footnotesize
\caption{X-ray Properties of the Targeted Radio Sources} 
\begin{center}
\begin{tabular}{ccccccc}
\hline\hline 
Name & $N_{\rm H}$ & Model & Parameter & $F_{2-8 {\rm keV}}$ $^{\rm a}$  & $\chi^{2}$/d.o.f. & Variability$^{\rm b}$\\ 
     & [$10^{20}$\,cm$^{-2}$] & &  &  [erg\,cm$^{-2}$\,s$^{-1}$] & & $\chi^{2}$/d.o.f. \\ 
\hline
NVSS~J092357+150518 & 3.51 (fixed) & PL & $\gamma$=2.0 (fixed)$^{c}$ & $<$ 1.37 $\times$ 10$^{-14}$ & $-$ & $-$\\
NVSS~J150229+555204 & 1.46 (fixed) & PL & $\gamma$=1.8$^{+0.3}_{-0.2}$ & 4.3$^{+1.1}_{-1.0}$ $\times$ 10$^{-14}$ & 29.5/32 & 11.7/13 \\
\hline
\end{tabular}
\end{center}
Notes. The best fit models are presented with the best fit parameters. \\
$^{\rm a}$The Fifth column shows the unabsorbed X-ray flux in the 2$-$8 keV band.\\
$^{\rm b}$$\chi^{2}$ value calculated from a fit to the X-ray light curve assuming a constant count rate.\\
$^{\rm c}$$\gamma$ is the photon index, where dN/dE $\propto$ E$^{-\gamma}$\\
\label{tab:fits_target}
\end{table}

\begin{table}[m]
\small
\caption{Positions and Detection Significances of X-ray Sources Detected Inside the 2FGL Error Regions}
\begin{center}
\begin{tabular}{cccccc}
\hline\hline
2FGL Name & Source Number & \multicolumn{2}{c}{Position} & Significance & Extraction Radius\\  
 & & RA[deg] & DEC[deg] & & \\ 
\hline 
2FGL J0923.5+1508 & 1 & 140.969(2) & 15.070(1) & 15.1$\sigma$ & 2$'$ \\
                  & 2 & 140.896(2) & 15.129(1) & 8.5$\sigma$ & 2$'$\\ 
                  & 3 & 140.980(4) & 15.164(2) & 6.9$\sigma$ & 2$'$\\ 
                  & 4 & 140.944(6) & 15.006(1) & 13.4$\sigma$& 2$'$\\
2FGL J1502.1+5548 & 5 & 225.578(1) & 55.812(1) & 25.2$\sigma$& 2$'$\\
                  & 6 & 225.570(3) & 55.756(2) & 21.3$\sigma$& 1$'$\\
                  & 7 & 225.524(3) & 55.744(2) & 16.0$\sigma$& 1$'$\\
                  & 8$^{a}$ & 225.387 & 55.849 & 24.6$\sigma$& 2$'$\\
\hline 
\end{tabular} 
\end{center} 
\label{tab:det_srcs}
$^{a}$Since source 8 is a diffuse source, the listed position is the center of the extraction region of the source events
\end{table}

\begin{table}[m]
\footnotesize
\caption{Spectral Parameters of Other Detected Sources} 
\begin{center}
\begin{tabular}{cclclccc} 
\hline\hline 
\multicolumn{2}{c}{Name} & $N_{\rm H}$ & Model & Parameter & F$_{2-8 {\rm keV}}$  & $\chi^{2}$/d.o.f.\\ 
 & & [$10^{20}$\,cm$^{-2}$] & &  &  [erg\,cm$^{-2}$\,s$^{-1}$] & \\ 
\hline
2FGL J0923.5+1508 & 1 & 3.51 (fixed) & PL & $\gamma$=2.0$^{+0.2}_{-0.2}$ & 4.3$^{+0.9}_{-0.8}$ $\times$ 10$^{-14}$ & 32.3/27\\
                  & 2 & 3.51 (fixed) & PL & $\gamma$=1.4$^{+0.4}_{-0.4}$ & 3.8$^{+1.4}_{-1.3}$ $\times$ 10$^{-14}$ & 15.2/14\\
                  & 3$^{a}$ & 5.0$^{+8.2}_{-3.0} \times 10^{2}$ & PL & $\gamma$=2.0 (fixed) & 5.5$^{+4.2}_{-2.7}$ $\times$ 10$^{-14}$ & 4.72/6\\
                  & 4 & 3.51 (fixed) & PL & $\gamma$=2.0$^{+0.3}_{-0.3}$ & 2.6$^{+0.8}_{-0.7}$ $\times$ 10$^{-14}$ & 14.4/20\\
2FGL J1502.1+5548 & 5 & 1.46 (fixed) & APEC+PL & $kT=1.2^{+0.1}_{-0.2}$ keV & 9.2$^{+0.9}_{-1.1}$ $\times$ 10$^{-14}$ & 107.9/95\\
                  &   &              &         & $\gamma$=2.3$^{+0.1}_{-0.1}$ & &  \\
                  & 6 & 1.46 (fixed) & PL & $\gamma$=1.8$^{+0.1}_{-0.1}$ & 1.1$^{+0.2}_{-0.2}$ $\times$ 10$^{-13}$ & 26.1/28\\
                  & 7 & 47$^{+45}_{-32}$ (fixed) & PL & $\gamma$=1.7$^{+0.4}_{-0.4}$ & 1.6$^{+0.3}_{-0.3}$ $\times$ 10$^{-14}$ & 7.3/16\\
                  & 8$^{b}$ & 12$^{+11}_{-9}$ & PL & $\gamma$=2.3$^{+0.3}_{-0.3}$ & 8.7$^{+1.6}_{-1.6}$ $\times$ 10$^{-12}$ & 108.6/77\\
\hline
\end{tabular}
\end{center}
Notes. The best fit models are presented with the best fit parameters. The fifth column shows the unabsorbed X-ray flux in the 2$-$8 keV band. $N_{\rm H}$ values derived from \citet{dic90} are 3.51 $\times 10^{20}$ cm$^{-2}$ and 1.46 $\times 10^{20}$ cm$^{-2}$ for the direction of 2FGL~J0923.5+1508 and 2FGL~J1502.1+5548, respectively.\\
$^{a}$This source is affected by extreme absorption at lower energy $E < 1$ keV and is significantly detected only with the XIS0 and XIS3 detectors. Therefore, the parameters are determined using the data obtained by these two detectors.\\
$^{b}$This source overlaps with the damaged part of the XIS0 CCD chip events that are discarded. Therefore, we used X-ray events of only XIS1 and XIS3 for the analysis.
\label{tab:fits}
\end{table}

\begin{table}[m]
\small
\caption{Infrared and Optical Counterparts of the Detected X-ray Sources}
\begin{center}
\begin{tabular}{ccl}
\hline\hline
Source Number & Counterpart Name & Value$^{a}$ \\ 
\hline 
1 & WISE J092351.81+150409.3 & w1 = 15.67, w2 = 14.62, w3 = 12.27, and w4 = 8.46\\
  & SDSS~J092351.82+150409.2 & u = 20.54, g = 20.10, r = 20.09, i = 20.22, and z = 20.21\\
  & SDSS~J092353.10+150416.4$^{b}$ & u = 23.76, g = 22.57, r = 21.70, i = 21.46, and z = 21.39\\
2 & WISE~J092333.95+150750.4 & w1 = 17.35, w2 = 15.81, w3 = 11.83, and w4 = 8.92\\ 
  & SDSS~J092333.94+150750.0 & u = 21.09, g = 20.95, r = 20.64, i = 20.24, and z = 20.47\\
3 & WISE~J092355.87+150949.2 & w1 = 17.24, w2 = 16.89, w3 = 11.90, and w4 = 8.59\\ 
  & SDSS~J092356.04+150951.2 & u = 24.52, g = 22.54, r = 21.03, i = 20.33, and z = 19.86\\
4 & SDSS~J092346.64+150026.8 & u $>$ 22.3, g $>$ 23.3, r = 22.28, i = 21.58, and z $>$ 20.8$^{c}$\\
5 & WISE~J150218.48+554830.9 & w1 = 9.83, w2 = 9.86, w3 = 9.80, and w4 = 8.96\\
  & SDSS~J150218.50+554830.8 & u = 13.52, g = 12.16, r = 11.68, i = 11.48, and z = 11.40\\
6 & WISE~J150216.01+554520.8 & w1 = 16.75, w2 = 16.28, w3 = 12.88, and w4 = 8.98\\ 
  & SDSS~J150216.03+554521.3 & u = 22.68, g = 22.11, r = 20.80, i = 20.27, and z = 19.80\\
7 & FIRST~J150205.3+554412 & 7.89 mJy\\ 
8 & MHL~J150137.1+555056$^{d}$ & r = 18.53\\ 
\hline 
\end{tabular} 
\end{center} 
\label{tab:counterparts}
$^{a}$Units of column three are magnitudes for infrared and optical counterparts and flux density for radio counterparts\\
$^{b}$The nearest optical counterpart without infrared counterpart\\
$^{c}$Upper limits of SDSS sources are 5$\sigma$ detection limits\\
$^{d}$Quoted from \citet{wen12}
\end{table}

%% file: figures_v2.tex
\begin{figure}[m] 
\begin{center}
\includegraphics[width=150mm]{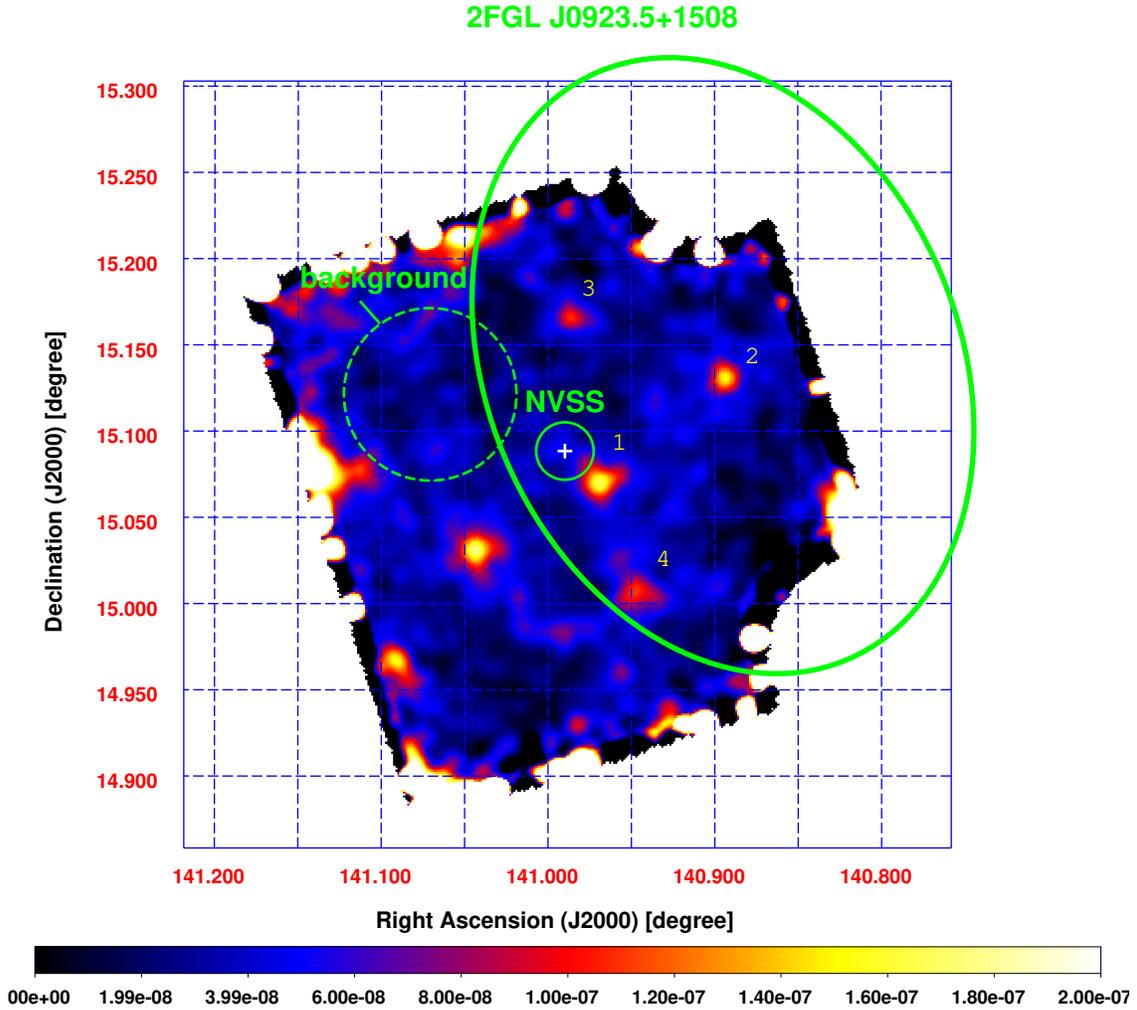}
\end{center}
\caption{X-ray image of 2FGL~J0923.5+1508 obtained by {\it Suzaku}/XIS0+3 (FI CCDs) in the 0.5-8 keV energy band. Thick solid ellipse denotes the 95\% positional error of 2FGL~J0923.5+1508. Thin solid and dashed circles show the source and background regions, respectively. A white cross indicates the radio position of NVSS~J092357+150518.} 
\label{fig:imgJ0923} 
\end{figure} 

\begin{figure}[m] 
\begin{center} 
\includegraphics[width=150mm]{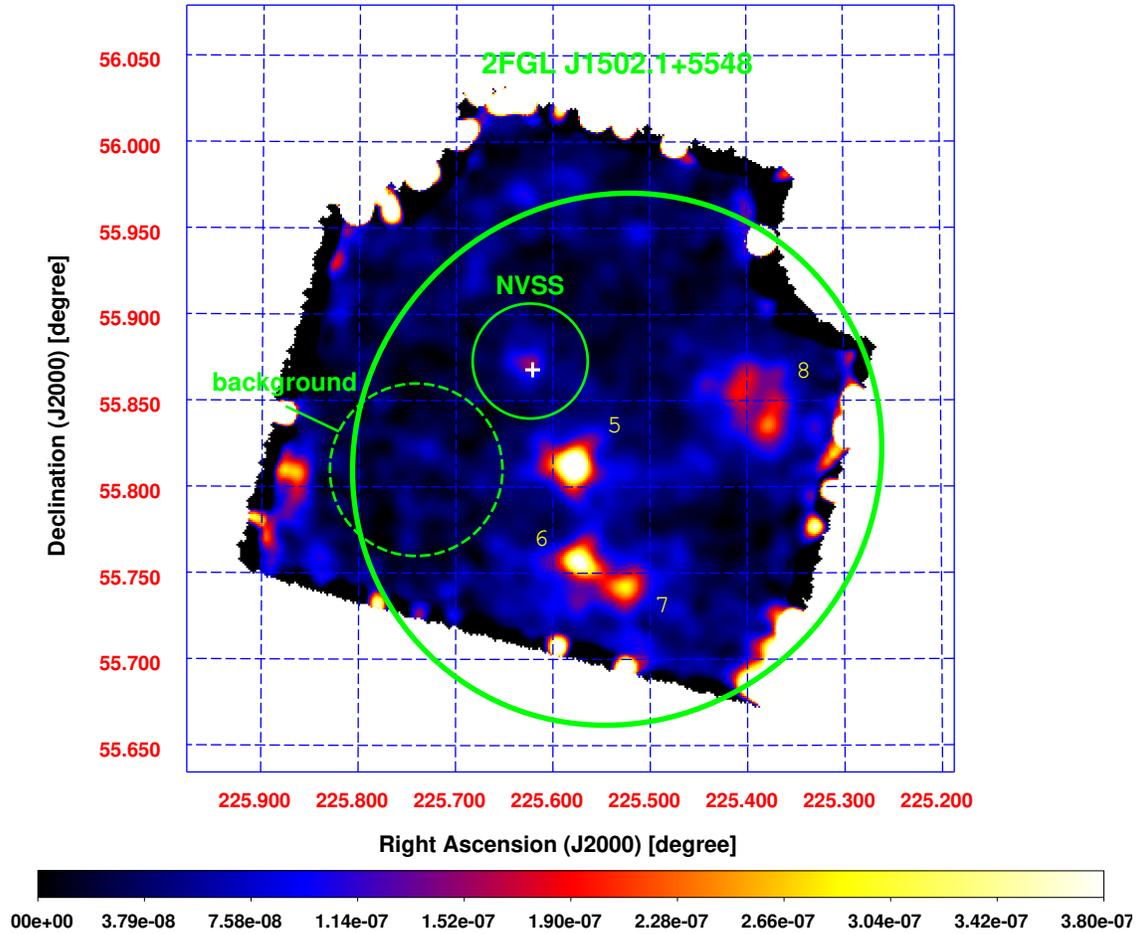} 
\end{center} 
\caption{X-ray image of 2FGL~J1502.1+5548 obtained by {\it Suzaku}/XIS0+3 (FI CCDs) in the 0.5-8 keV energy band. Thick solid ellipse denotes the 95\% positional error of 2FGL~J1502.1+5548. Thin solid and dashed circles show the source and background regions, respectively. A white cross indicates the radio position of NVSS~J150229+555204.} 
\label{fig:imgJ1502} 
\end{figure} 

\begin{figure}[m] 
    \begin{center}
      \includegraphics[width=80mm,angle=-90]{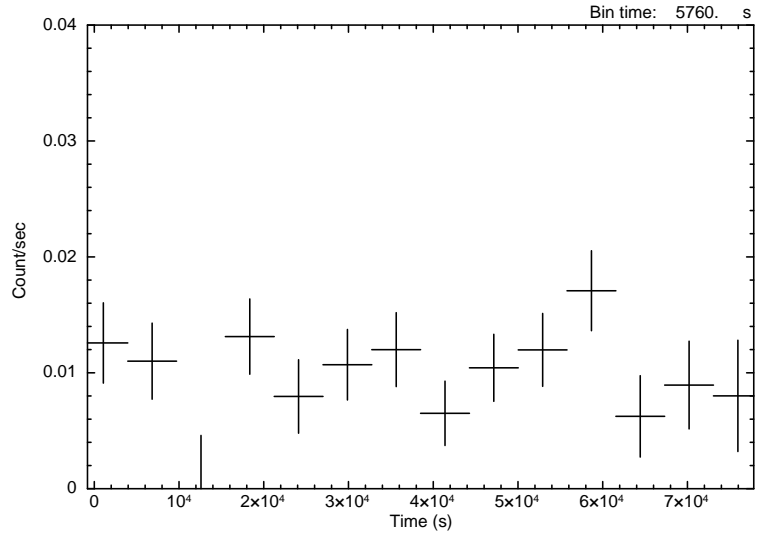}
    \end{center}
  \caption{{\it Suzaku}/XIS light curve of the X-ray counterpart of NVSS~J150229+555204.} 
  \label{fig:lcJ1502} 
\end{figure}

\begin{figure}[m] 
    \begin{center}
      \includegraphics[width=80mm,angle=-90]{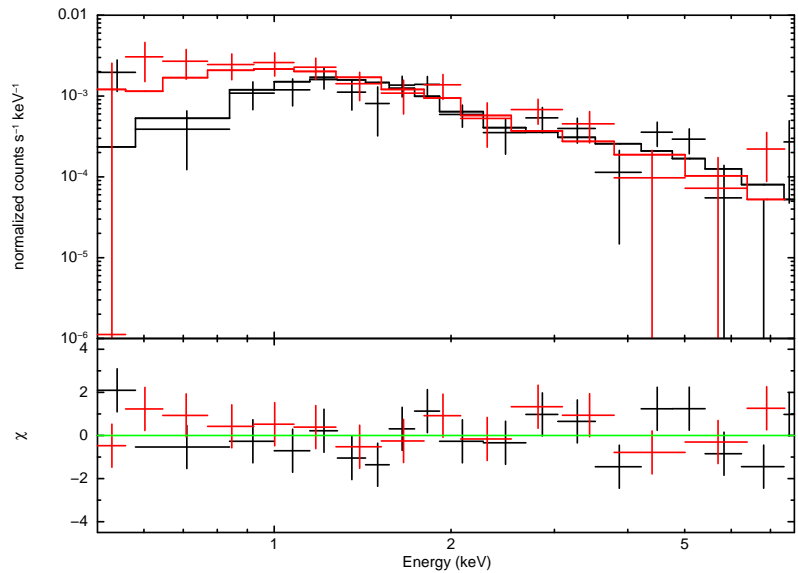}
    \end{center}
\caption{{\it Suzaku}/XIS spectrum of the X-ray counterpart of NVSS~J150229+555204 fitted with an absorbed power-law model. Black plots show the FI data and red plots show the BI data. Black solid and red solid curves are the best fit models of FI and BI data, respectively.} 
\label{fig:specJ1502} 
\end{figure} 

\begin{figure}[m] 
\begin{center} 
\includegraphics[width=150mm]{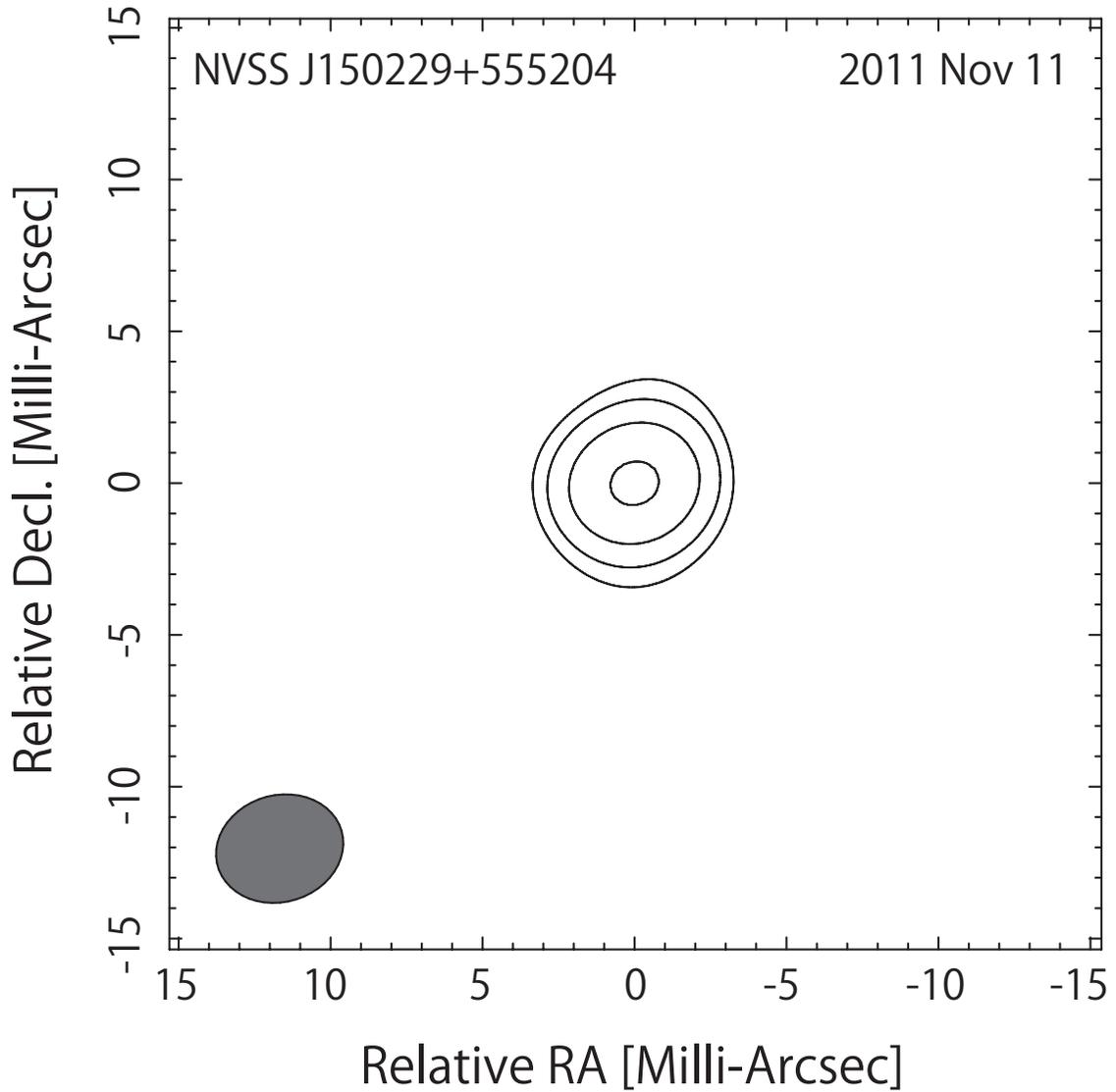} 
\end{center} 
\caption{VLBI image of NVSS J150229+555204 with natural weighting at 6.7 GHz. The epoch is indicated on the top of the panel as "YYYY MMM DD". The first contour intensity is 4.1 mJy beam$^{-1}$, which corresponds to three times the image noise level, and contour levels increase by a factor of 2. Also, the restoring beam size is indicated in the bottom-left corner.} 
\label{fig:vera_j1502} 
\end{figure} 

\begin{figure}[m] 
  \begin{minipage}{\hsize}
    \begin{center}
      \subfigure[2FGL~J0923.5+1508]{\includegraphics[width=120mm]{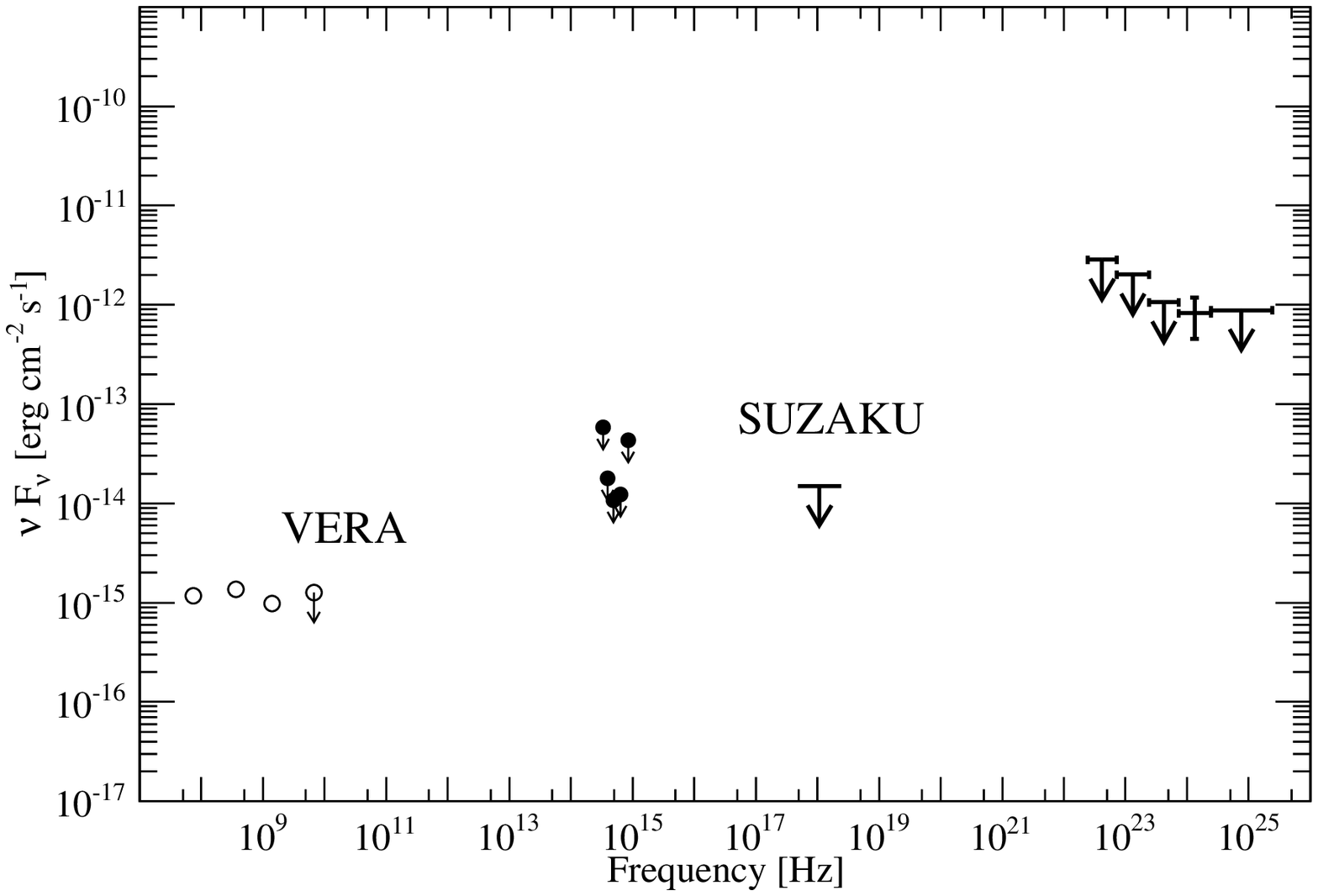}}
    \end{center}
  \end{minipage}
  \begin{minipage}{\hsize}
    \begin{center}
      \subfigure[2FGL~J1502.1+5548]{\includegraphics[width=120mm]{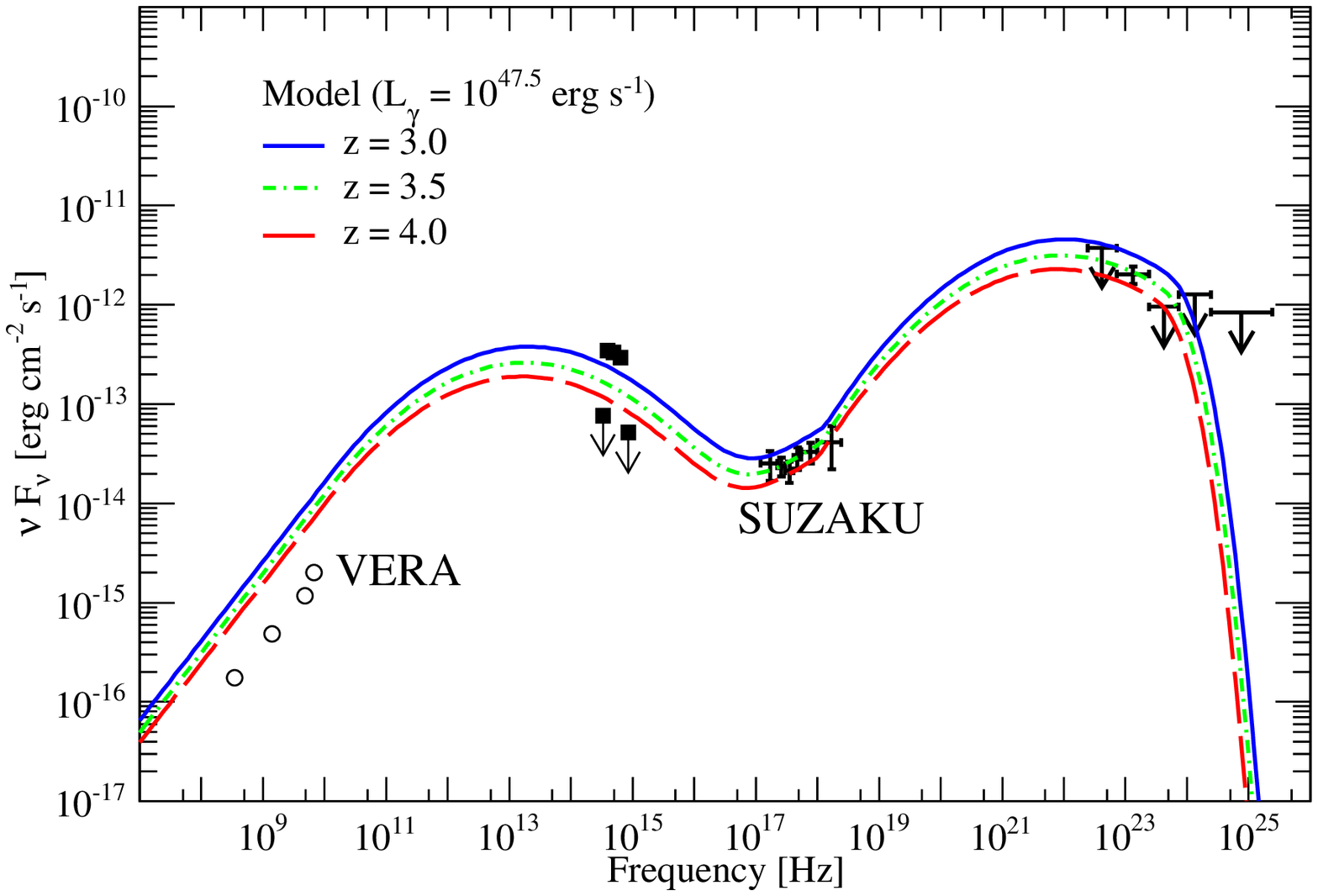}}
    \end{center}
  \end{minipage}
  \caption{SEDs of the gamma-ray sources when assuming the targeted radio sources as the counterparts. Gamma-ray data points are taken from the 2FGL catalog \citep{2FGL}. The radio data points are adopted from selected catalogs listed in the notes of Table~\ref{tab:radio_ctp}, and our newly conducted VERA observations. The optical data are taken from the SDSS catalog \citep{ade11}. The optical upper limits represent 5$\sigma$ detection limits for each color band in the SDSS. The blazar SED sequence model of \citep{ino09} accounting for intergalactic attenuation with the EBL model of \citep{ino12} are overlapped to the SED of 2FGL~J1502.1+5548. The red dashed , green dot-dashed, and the blue solid curves represent the blazar sequence models with gamma-ray luminosity $L_{\gamma} = 10^{47.5}$ erg s$^{-1}$ assuming redshifts z = 3.0, 3.5, 4.0, respectively.}
  \label{fig:seds}
\end{figure}

\begin{figure}[m] 
\begin{center}
\includegraphics[width=150mm]{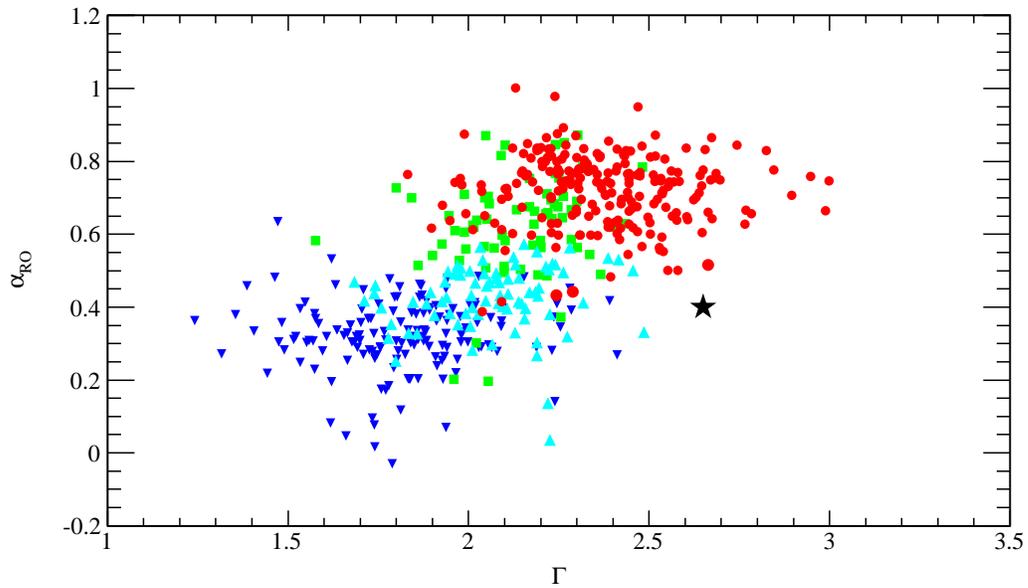}
\end{center}
\caption{Radio to optical rest-frame broad-band spectral indices versus gamma-ray photon indices of blazars listed in the 2LAC \citep{ack11}. Dark blue inverted triangles: HSP BL Lacs, light blue triangles: ISP BL Lacs, green squares: LSP BL Lacs, red circles: FSRQs. All the values are taken from 2LAC. We also plot the parameters of 2FGL~J1502.1+5548 with a black star. The redshift of 2FGL~J1502.1+5548 is assumed to be 3.5.}
\label{fig:a-gpl}
\end{figure}